\documentstyle[12pt]{article}
\def\beq{\begin{equation}}
\def\eeq{\end{equation}}
\def\bea{\begin{eqnarray}}
\def\eea{\end{eqnarray}}
\def\bem{\begin{math}}
\def\eem{\end{math}}
\def\bit{\begin{itemize}}
\def\eit{\end{itemize}}
\def\bla{\begin{flushright}}
\def\ela{\end{flushright}}
\def\qq2{$Q^2$}               
\def\aa1{$A_1(x,Q^2)$}        
\def\ff1{$F_1(x,Q^2)$}        
\def\gg1{$g_1(x,Q^2)$}        
\begin{document}
\vskip 3cm
\begin{center}
{\Large{\bf
The $Q^2$ dependence of the measured asymmetry $A_1$ \\
from the similarity of $g_1(x,Q^2)$ and  $F_3(x,Q^2)$
structure functions 
}}
\end{center}
\vskip 1.7cm
\begin{center}
{\bf A.V.Kotikov}
\footnote{
~E-mail: kotikov@sunse.jinr.ru; Anatoli.Kotikov@cern.ch}
and {\bf D.V.Peshekhonov}
\footnote{
~E-mail: Dimitri.Pechekhonov@infn.trieste.it;
Dmitri.Pechekhonov@cern.ch\\
~presently at INFN sezione di Trieste, Italia}
\\
\vskip 0.5cm

{\it Particle Physics Laboratory\\ Joint Institute for Nuclear Research\\ 
141980 Dubna, Russia.}
\end{center}
\vskip 3cm
{\large{\bf Abstract}}\\
We propose a new approach for taking into account the $Q^2$ dependence of
measured asymmetry $A_1$.
This approach is based on the similarity of the $Q^2$ behaviour 
and the shape of 
the spin-dependent structure function $g_1(x,Q^2)$ and spin averaged 
structure function $F_3(x,Q^2)$. 
The analysis is applied on 
available experimental data.
\vskip .5cm \hskip -.56cm
PACS number(s): 13.60.Hb, 11.55.Hx, 13.88.+e \\
\newpage
\section{Introduction}

In a recent years there has been a significant progress in the study 
of the spin-dependent structure function (SF)
 $g_1(x,Q^2)$ (see \cite{SMCp}-\cite{Q2E154}).
The direct measurement of SF $g_1$ is very elaborate procedure (see, however,
\cite{Gagu})
and ordinary its value is extracted from the  spin dependent 
asymmetry $A_1$ (see, for example \cite{AEL, Ram}) in agreement with the 
following formula:
\begin{eqnarray}
g_1(x,Q^2)~=~A_1(x,Q^2) \cdot F_1(x,Q^2),
\label{a}
\end{eqnarray}
where $F_1(x,Q^2)$~is the spin average SF.\\
\hskip -.56cm
The asymmetry $A_1(x,Q^2)$ is
closely connected with the ratio of
polarized and unpolarized cross-sections and may be "easy"
measured due to cancelation of many experimental uncertantities.
Experimentally asymmetry is extracting only at  few points 
$Q^2_{1i}, ...,Q^2_{ni}$ for each $x_i$ bin. To study the properties of
$g_1(x,Q^2)$ and to calculate the values of  spin dependent
sum rules \cite{Bj,EJ} we have
to know the $A_1$ as a function of $Q^2$.\\ 
\hskip -.56cm
The most popular
assumption applied on $A_1$ \cite{EK93} is 
\begin{eqnarray}
A_1(x,Q^2) = A_1(x), \label{a1}
\end{eqnarray}
It means that SF $g_1$ and  $F_1$ have the 
same $Q^2$ dependences 
what does not  following from the theory. 
On the contrary, the 
behaviour of $F_1$ and $g_1$ as a functions of $Q^2$ is expected to be 
different due to the difference between polarized and unpolarized splitting 
functions\footnote{except the leading order of the quark-quark interaction.}.\\
There are several approaches \cite{Q2A}-\cite{NST} how to take into account 
the $Q^2$ dependence of $A_1$.
They are 
based on the different approximate solutions of the DGLAP equations.
Some of them have been used already by Spin Muon Collaboration (SMC)
and E154 Collaboration in the last analyses of experimental data
(see \cite{SMC} and \cite{Q2E154}, respectively). These
approaches \cite{Q2A}-\cite{NST} lead to similar results in 
$g_1(x,Q^2)$\footnote{ The form of the $Q^2$-dependence
for $A_1$ is different in approaches \cite{Q2A}-\cite{NST}. However, all of
them are in agreement with the weak $Q^2$-dependence at moderate $x$ 
and quite strong at small $x$ region.
} which are contrast with the calculations
based on Eq.(\ref{a1}).\\

In this article we suggest another method how to take into account $Q^2$
dependence of $A_1$
which is based on the 
observation that the splitting functions of  the DGLAP equations for
the SF $g_1$ and  $F_3$ and  the shapes of these SF 
are similar in a wide $x$ range.
Our approach allows to set  $Q^2$-dependence of $A_1$ in a very simple way 
(see Eq.(\ref{5})) and leads to the results
(some of them have been recently presented \cite{DIS98} on the Workshop
DIS98), which are similar 
with ones based on the  DGLAP evolution.\\

\section{The $Q^2$ dependence of 
structure functions}

Lets consider the $Q^2$ evolution of 
nonsinglet (NS) and singlet (SI) parts of the SF separately.\\

\hskip -.56cm
For the SF $F_3$ and
the nonsinglet (NS) parts of $g_1$ and $F_1$
the corresponding DGLAP equations can be presented
as\footnote{We use $\alpha(Q^2)= \alpha_s(Q^2)/{4 \pi}$.} as:\\
\bea
{dg_1^{NS}(x,Q^2) \over dlnQ^2} &=& -{1 \over 2} \gamma_{NS}^-(x, \alpha)
\otimes g_1^{NS}(x,Q^2),
\nonumber \\
{dF_1^{NS}(x,Q^2) \over dlnQ^2} &=& -{1 \over 2} \gamma_{NS}^+(x, \alpha)
\otimes F_1^{NS}(x,Q^2),
\label{1} \\
{dF_3(x,Q^2) \over dlnQ^2} &=& -{1 \over 2} \gamma_{NS}^-(x, \alpha)
\otimes F_3(x,Q^2),
\nonumber
\eea
where the symbol $\otimes$ means the Mellin convolution:
\bea
f_1(x) \otimes f_2(x) \equiv \int^1_x \frac{dz}{z} f_1(z) f_2(\frac{x}{z})
\nonumber
\eea
The splitting functions $\gamma^{\pm}_{NS}(x,\alpha)$ are the reverse
Mellin transforms of the anomalous dimensions $\gamma^{\pm}_{NS}(n, \alpha)= \alpha \gamma^{(0)}_{NS}(n) + \alpha^2 \gamma^{\pm (1)}_{NS}(n) + O(\alpha^3)$ and the Wilson coefficients\footnote{We consider here structure functions 
but not the parton distributions. Note also that $b^{+}_{NS}(n)$ and
$b^{-}_{NS}(n)$ can be defined as $b_{1,NS}(n) = b_{2,NS}(n) - b_{L,NS}(n)$ and
$b_{3,NS}(n)$.}
$\alpha b^{\pm}(n) + O(\alpha^2)$:
 \begin{eqnarray}
\gamma^{ \pm}_{NS}(x,\alpha) ~=~ \alpha
  \gamma^{(0)}_{NS}(x) + \alpha^2 \biggl(
  \gamma^{\pm (1)}_{NS}(x) +
  2\beta_0 b^{\pm}(x)  \biggr) + O(\alpha ^3),
\label{2}  \end{eqnarray}
where $\beta (\alpha)= - \alpha^2 \beta_0 -
 \alpha^3 \beta_1 + O(\alpha^4)$ is QCD $\beta$-function.\\
Eqs. (\ref{1}) show that the 
DGLAP equations for $F_3$ and for the NS part
of $g_1$  are the same
(it was obtained exactly
in first two orders of the perturbative QCD\footnote{It is easy
to demonstrate 
in any order of perturbation theory. The
SF $g_1$ and $F_3$ are the results of the $\gamma_5$ matrix
contribution to the lepton
and hadron parts of deep-inelastic
cross-sections, respectively. In the NS case 
there is only one $\gamma$-matrix trace,
connecting the lepton and hadron parts. Its contribution $\sim ~
tr(\gamma_5 \gamma_{\mu} \gamma_{\nu} \gamma_{\alpha} \gamma_{\beta} ....)$
is the same in both cases above. For the SI part of $g_1$ there are
diagrams with several traces, which arise in the second order of 
perturbation QCD and lead to the difference between the splitting
functions of SF $F_3$ and the SI part of SF $g_1$.}  
 \cite{Kodaira})
and differ from the one for $F_1$ already in the first subleading order
($\gamma^{+(1)}_{NS} \neq \gamma^{-(1)}_{NS}$ \cite{RoSa} and
$b^+_{NS} - b^-_{NS} = (8/3)x(1-x)$).\\

\hskip -.56cm
For the SI parts of $g_1$ and $F_1$ evolution equations are :\\
\bea
{dg_1^S(x,Q^2) \over dlnQ^2} &=& -{1 \over 2} \biggl[
\gamma_{SS}^{*}(x, \alpha) \otimes g_1^S(x,Q^2) +
\gamma_{SG}^{*}(x, \alpha) \otimes \Delta G(x,Q^2) \biggr],
\nonumber
\\
{dF_1^S(x,Q^2) \over dlnQ^2} &=& -{1 \over 2} \biggl[
\gamma_{SS}(x, \alpha) \otimes F_1^S(x,Q^2) +
\gamma_{SG}(x, \alpha) \otimes G(x,Q^2) \biggr],
\label{4}
\eea
where the SI splitting functions $\gamma_{Si}(x, \alpha)$
($i=\{S,G\}$) are represented as
 \bea
\gamma_{SS}(x,\alpha) &=& \alpha
  \gamma^{(0)}_{SS}(x) + \alpha^2 \biggl(
  \gamma^{(1)}_{SS}(x) +  b_G(x) \otimes \gamma^{(0)}_{GS}(x) +
  2\beta_0 b_S(x)  \biggr)
+ O(\alpha ^3),
\nonumber  \\
\gamma_{SG}(x,\alpha) &=& \frac{e}{f} \biggl[ \alpha
  \gamma^{(0)}_{SG}(x) + \alpha^2 \biggl(
  \gamma^{(1)}_{SG}(x) + b_G(x) \otimes \bigl(
  \gamma^{(0)}_{GG}(x) - \gamma^{(0)}_{SS}(x) \bigr)
+ 2\beta_0 b_G(x)
            \nonumber \\
&+&  b_S(x) \otimes \gamma^{(0)}_{SG}(x)
\biggr)
  \biggl] + O(\alpha ^3)  \label{4.1} \eea
and $e = \sum_i^f e^2_i$ is
the sum of charge squares of $f$ active quarks.
Equations for the polarized singlet splitting functions
$\gamma_{SS}^{*}(x, \alpha)$ and $\gamma_{SG}^{*}(x, \alpha)$
are similar. They can be obtained from (\ref{4.1}) by replacing
$\gamma^{(0)}_{SG}(x) \to \gamma^{*(0)}_{SG}(x)$,
$\gamma^{(1)}_{Si}(x) \to \gamma^{*(1)}_{Si}(x)$ and $b_i(x) \to
b^*_i(x)$ ($i=\{S,G \}$).\\

\hskip -.56cm
Careful consideration of  the quark part of (\ref{4}) and (\ref{4.1})
shows that
the value $b^*_s(x)$ ($b_s(x)$) coincides with $b^-(x)$ ($b^+(x)$).
The difference between
$\gamma_{NS}^{-(1)}(x)$ and $\gamma_{SS}^{* (1)}(x) +  b^*_G(x) \otimes
\gamma^{(0)}_{GS}(x)$  is negligible because it does not contain
a power singularity at $x \to 0$ (i.e. a singularity
at $n \to 1$ in momentum space).
Moreover, this difference decreases as $O(1-x)$ at $x \to 1$
\cite{MeNe}
(contrary to this, the difference between
$\gamma_{SS}^{(1)}(x) +  b_G(x) \otimes \gamma^{(0)}_{GS}(x)$ and
$\gamma_{SS}^{* (1)}(x) +  b^*_G(x) \otimes \gamma^{(0)}_{GS}(x)$
contains the power singularity at $x \to 0$ (see for example
\cite{Kodaira})).
Thus, the DGLAP equations for $F_3$ and the SI part of $g_1$ have
a close splitting functions, which are essentially different from the
splitting functions of the SI part of $F_1$. \\
The quark part of SI SF $g_1$ itself 
contains two 
components:
valence and sea. Valence part does not connect with the
gluon and obeys the DGLAP equation similar to first equation of (\ref{1}).
The sea part obeys the first equation of (\ref{4}). This contribution
was not observed  experimentally yet.\\
The contribution of the gluon distribution in $g_1$
is not so important for the modern data 
(see \cite{GRSV,GS96, Ram, NST, GRam})
(in a contrary to  unpolarized case):  
data are described well for extremally 
different values of $\Delta G(x,Q^2)$ and even for different  sign.
Hence we will neglet this term  in our analysis.\\
Thus, 
the valence component seems to dominate in SI part of $g_1$
at the range of the present experimental data\footnote{To support
of this point of view, one can see also the recent analysis of
E154 Collaboration \cite{Q2E154}, where the contribution of sea $+$
gluon and valence parts are divided, studied and presented in Fig. 2
of \cite{Q2E154}.}
and it allows us to expect
a similarity of SI part of $g_1$ and SF $F_3$.\\
 
\hskip -.56cm
As we saw from above, 
the shapes and 
DGLAP equations for $g_1$ and $F_3$
are very close in NS and SI analyses\footnote{The similar
shapes of SF $F_3$ and  $g_1$ in the range of measured values of $x$ 
one can see also in the analysis  \cite{BoSo}.}
and
both of them differ from the
corresponding equations for $F_1$.
This
similarity leads to close $Q^2$ dependence of SF $g_1(x,Q^2)$ and
$F_3(x,Q^2)$ .\\

\hskip -.56cm
The similarity of $Q^2$ dependence 
of $F_3(x,Q^2)$ and  $g_1(x,Q^2)$
may be also supported by some arguments following from analysis at
$x \to 0$ \footnote{At $x \to 1$ the behaviour of $g_1$ and 
$F_3$ (and $F_1$, too) should be similar because it is governed by valence 
quark distributions.}. 
Although all existed data in polarized DIS (exclude  two first SMC points) 
are outside from the region $x \leq 10^{-2}$, the study
of  small $x$ asymptotics is 
important as for
the future data, as for an extrapolation of the present data to the small $x$
egion.\\
The similarity of the splitting functions of  SF $F_3$ and 
$g_1$
have been already demonstrated and, thus,
 we have to discuss now the shapes of  $F_3$ and 
 $g_1$ at small $x$.
It is well known that SF $F_3$ is governed at small $x$ by $\rho$-meson 
trajectory and, thus,
\bea
F_3(x) \sim x^{-1/2}
\label{d.1}
\eea
The $Q^2$-evolution does not change this behaviour. For SF $g_1$ the 
situation is not so clear. From Regge analysis \cite{Hei} the NS part of
$g_1$ is governed by $a_1$ trajectory, i.e.
$
g_1(x) \sim x^{-\alpha_P(a_1)}
$
with the intersept $\alpha_P(a_1)$ having values
$0 \geq \alpha_P(a_1) \geq -1/2$ (see \cite{ElKa1}). However,  
the BFKL inspired approach \cite{BERns} leads to
more singular behaviour for NS 
part: 
\bea
g_1^{NS}(x) \sim x^{-0.45},
\label{d.3}
\eea
 which is close to (\ref{d.1}). 
For SI
part of $g_1$ an information is very poor. The BFKL inspeared approach
\cite{BERs}
leads to the small $x$ behaviour $g_1^{SI}(x) \sim x^{-1}$, but really SI part
 of $g_1$\footnote{Correctly,  sea component of SI part of  $g_1$,
which dominates here if it has nonzero magnitude.}
was not observed at small $x$ yet. It is also connected  with the
fact that the deutron SF $g_1^{d}(x)$, which is close to the SI component, 
is comparable with zero at small $x$.\\ 
As a consequences, the shapes of  SF $F_3$ and the
NS part of SF $g_1$ seems to be close also at small $x$ (if BFKL 
approach is correct at least). 
The SI part of $g_1$ may has another shape but modern experimental
data do not allow to study it.\\

\hskip -.56cm
The analysis discussed above allows us to conclude that
the function $A_1^*$ defined as:
\bea
A_1^*(x) = {g_1(x,Q^2) \over F_3(x,Q^2)}
\label{4.2}
\eea
has to be practically $Q^2$ independent in the hole region of modern
experimental data \cite{SMCp}-\cite{Q2E154}.
\\

\hskip -.56cm
In agreement with Eq.(\ref{4.2}) measured asymmetry $A_1(x_i,Q^2_i)$ can be 
found at some value of $Q^2$ as:
\bea
A_1(x_i,Q^2) =  {F_3(x_i,Q^2) \over F_3(x_i,Q^2_i)} \cdot
{F_1(x_i,Q^2_i) \over F_1(x_i,Q^2)} \cdot A_1(x_i,Q^2_i)
\label{5}
\eea

\section{Calculation of $A_1(x,Q^2)$ and $\Gamma_1(Q^2)$.}

To apply this approach we used\footnote{Similar analysis of SMC data of
\cite{SMCp} has been done in  \cite{KP} using
old CCFR data \cite{CCFR}.} the 
SMC \cite{SMC}, E143 \cite{E143} and  E154 \cite{E154n,Q2E154}
collaboration data.
To use relation (\ref{5}) we have parametrized CCFR data on 
$F_2(x,Q^2)$ and $xF_3(x,Q^2)$ 
\cite{CCFRN} in the form the same with NMC fit of the structure function 
$F_2(x,Q^2)$ \cite{NMC} (see Appendix). 
To obtain structure function $F_1(x,Q^2)$ we take the parametrization of 
the CCFR data on $F_2(x,Q^2)$ \cite{CCFRN} and SLAC parametrization of 
$R(x,Q^2)$ \cite{SLAC} 
and use relation :
\begin{equation}
F_1(x,Q^2)= \frac{F_2(x,Q^2)}{2x(1+R(x,Q^2))} \cdot 
(1+ \frac{4M^2x^2}{Q^2}) ,
\label{5.1}
\end{equation}
where  $M$ is the proton mass. \\
The fact that we use in Eq.(\ref{5}) parametrizations of CCFR data
\cite{CCFRN} for both 
SF $xF_3(x,Q^2)$ and $F_2(x,Q^2)$ allows to avoid systematical uncertanties and
nucleon correlation in nuclei.\\
Fig. 1 shows the ratio $A_1(Q^2)/A_1(5GeV^2)$ obtained with Eq.(\ref{5}).
Comparison of Fig. 1 with the results of E154 Collaboration (Fig. 4 in 
\cite{Q2E154}) shows quite good agreement.\\
 \hskip -.56cm
The SF $g_1(x,Q^2)$ was evaluated
using Eq.(\ref{a}) 
where spin average SF $F_1$ has been calculated using NMC 
parametrization of $F_2(x,Q^2)$ \cite{NMC}.  
Results are presented in Fig. 2 and Fig. 3 for E154 and SMC data,
respectively. Our results are in excelent agreement with the
SMC and E154 Collaboration analyses  based 
on direct DGLAP evolution
(see \cite{SMC} and \cite{Q2E154}, respectively).\\

%
%

\hskip -.56cm
To make a comparison with the theory predictions on the sum rules
we have calculated also the 
first moment
value of the structure function $g_1$ at different $Q^2$
\begin{equation}
\Gamma_1= \tilde \Gamma_1 + \Delta \tilde \Gamma_1,  \nonumber
\end{equation} 
where 
\begin{equation}
\tilde \Gamma_1(Q^2)=\int_{x_{min}}^{x_{max}} g_1(x,Q^2)dx
~~\mbox{ and }~~
\Delta\tilde \Gamma_1(Q^2)=\int^{x_{min}}_{0} g_1(x,Q^2)dx +
\int^{1}_{x_{max}} g_1(x,Q^2)dx
\nonumber \end{equation} 
are
the integral through  the measured kinematical $x$ region plus an
 estimation for unmeasured ranges, respectively.\\ 
The value of $\Delta\tilde \Gamma_1$ coming from the unmeasured 
$x$-regions was estimated using original
methodics by 
``owner-collaborations''. 
%
We have to note here that the methodic 
of $\Delta\tilde \Gamma_1$ estimation 
may leads to some underestimation of 
$g_1^{p,d,n}(x,Q^2)$ at small $x$ and of $ \Gamma ^{p,d,n}(Q^2)$, 
as a consequense (see the careful analysis in \cite{GRSV}). 
To clear up this situation it is necessary to have
more precise data at small $x$.\\
Values of $\Gamma_1(Q^2)$ which are obtained from the exact solution of the 
DGLAP evolution equation \cite{SMC,Q2E154} of $g_1(x,Q^2)$ and $A_1$
\cite{SMCp}-\cite{Q2E154} and in our approach on the scaling of $A_1^*$ are
quite close each other for all cases discussed here. 
Thus all approaches lead to the similar conclusions on $ \Gamma ^{p,d,n}(Q^2)$
and the results are in a strong disagreement with  the theoretical predictions
\cite{LRV1}, we will consider the effect of $A_1^*$ scaling only for the 
Bjorken Sum rule $\Gamma_1^p -\Gamma_1^n$.\\
%
%
%
Deuteron SMC and E143 data allow us to extract the value of $\Gamma_1^n$: 
\begin{equation}
\Gamma_1^p + \Gamma_1^n=\frac{2\Gamma_1^d}{1-1.5w_d},  \nonumber
\end{equation}
where $w_d$=0.05 is the the probability of the deutron 
to be in a D-state.
Knowledge of proton and neuteron first momenta $\Gamma_1^{p,n}$ allows 
to test the Bjorken sum rule:
\begin{equation}
\Gamma_1^{p-n} \equiv  \int \limits_{0}^{1}(g_1^p(x,Q^2)-g_1^n(x,Q^2))dx=
\Gamma_1^p -\Gamma_1^n   \nonumber
\end{equation}
Results
are presented in the Table 1 in comparison with 
values published by SMC, E143 and E154 Colaborations and with the theoretical
predictions computed in the third order in the QCD $\alpha_s$ \cite{LV}.\\
Let us now describe the main results, which follow from the Table 1 and
 the  Figures.
\begin{itemize}
\item Our description of the $Q^2$ evolution of the asymmetry \aa1 
has very simple form (\ref{5}) but gives results which
are in good agreement with 
a powerfull analyses \cite{GRSV,GS96,Q2E154}.
\item
Results on $g_1(x,Q^2)$ are in excelent agreement with  SMC and
E154 Collaborations analyses, based 
on direct DGLAP evolution.
\item Our method allows to 
test 
sum rules in a simple way with a good accuracy.
Obtained results on the $\Gamma_1^p - \Gamma_1^n$ show that used experimental 
data 
well confirm the Bjorken sum rule  prediction.
\end{itemize}
\hskip -.56cm

\section{Conclusion} 
We have considered the $Q^2$ evolution of the asymmetry \aa1
based on the similarity of $Q^2$ dependence of the SF $g_1(x,Q^2)$ and 
$F_3(x,Q^2)$\footnote{The usefull parametrizations of SF
$F_2(x,Q^2)$ and $xF_3(x,Q^2)$ are obtained for new CCFR data and
presented in Appendix.}.
Obtained results on  $g_1(x,Q^2)$
are in excelent agreement with the corresponding results of SMC and
E154 Collaborations, based 
on  direct DGLAP evolution.
Our test of the Ellis-Jaffe sum rules for the proton, deuteron and neutron 
are very close to the values published by Spin Muon, E143 and E154 
Collaborations. 
However, the 
corrections coming due to $Q^2$ evolution of asymmetry \aa1 have an opposite 
signes for the proton and deutron. It leads to the 
improvement in agreement between the experiment and the 
theoretical prediction on the Bjorken sum.\\
%
We believe that future precise data will illuminate a violation of 
our hypothese (probably, at very small $x$ values: $x \leq 10^{-3}$).
This violation will indicate clearly the appearence of nonzero 
contributions from
sea quark and gluon components of SF $g_1(x,Q^2)$ having quite singular 
 shapes at small $x$\footnote{In the case of a similar shapes of SI and NS 
components the separation and study of them will be
quite elaborate procedure.} (see the carefull analysis in \cite{HNS}). 
Thus, check of the 
$Q^2$-dependence of the ratio $A_1^* = g_1/F_3$  for a future
precise data leads to make a qualitative estimations
shapes and 
$Q^2$-dependences of gluon and sea quark
distributions.\\

\vskip 0.5 cm

{\large \bf Acknowledgements}\\

\hskip -.56cm
We are grateful to W.G.~Seligman for
providing us the available CCFR data of Refs.\cite{CCFRN, CCFR},
to A.V.~Efremov and O.V. Teryaev for interest to this work and discussions
and to A.P. Nagaitsev for a collaboration in the beginning
of this study.\\

\hskip -.56cm

\hskip -.56cm
{\large \bf Appendix}\\

\hskip -.56cm
The parametrizations are used for CCFR data \cite{CCFR} :
\bea
xF_3(x,Q^2) = F_3^a  \cdot { \biggl( {
log(Q^2/\Lambda ^2) \over log(Q^2_0/\Lambda ^2) } \biggl) }^
{F_3^b}
~\mbox{ and }~
F_2(x,Q^2) = F_2^a  \cdot { \biggl( {
log(Q^2/\Lambda ^2) \over log(Q^2_0/\Lambda ^2) } \biggl) }^
{F_2^b}
,
\nonumber
\eea
where
\bea
F_3^a &=& x^{C_1} \cdot (1-x)^{C_2} \cdot
\biggl( C_3+C_4 \cdot (1-x) +C_5 \cdot
(1-x)^2
\nonumber \\ & &~+~
C_6 \cdot (1-x)^3 +C_7 \cdot (1-x)^4 \biggr) 
\nonumber \\ & & \nonumber \\
F_2^a &=& x^{B_1} \cdot (1-x)^{B_2} \cdot
\biggl( B_3+B_4 \cdot (1-x) +B_5 \cdot
(1-x)^2
\nonumber \\ & &~+~
B_6 \cdot (1-x)^3 +B_7 \cdot (1-x)^4 \biggr) 
\nonumber \\ & & \nonumber \\
F_3^b &=& C_{8}+C_{9} \cdot x+{C_{10} \over x+C_{11}}
~\mbox{ and }~
F_2^b = B_{8}+B_{9} \cdot x+{B_{10} \over x+B_{11}}
\nonumber 
\eea
and $Q^2_0 = 20~ {\rm GeV}^2$, $\Lambda = 337~{\rm MeV}$. The values of
$Q^2_0$ and $\Lambda $ are fixed in agreement with CCFR analysis 
\cite{CCFRN}.
The values of the coefficients $C_i$ ($i=1,...,11$) 
and $B_i$ ($i=1,...,15$) are 
given in Table 2.\\


%

%
%
\vskip 2cm
\newpage
{\Large {\bf Figure Captions}}\\
\vskip 1cm
\begin{description}
\item [Figure 1.]  $Q^2$ dependence of the ratio
 $A_1(x,Q^2)/A_1(x,5GeV^2)$.~ 
\item [Figure 2.]
Structure function $xg^n_1(x,Q^2)$ evolved to $Q^2 = 5 GeV^2$
using our eq.(\ref{5}); DGLAP NLO evolution; the assumption that
$g^n_1/F^n_1$ is independent of $Q^2$. Last two sets are taken
from \cite{Q2E154}.
\item [Figure 3.]
Structure function $xg^p_1(x,Q^2)$ evolved to $Q^2 = 10 GeV^2$
using our eq.(\ref{5}); the assumption that
$g^n_1/F^n_1$ is independent of $Q^2$; DGLAP NLO evolution according
to the analyses of \cite{Q2A,GRSV}. Last two sets are taken from \cite{SMC}.

\end{description}

\vskip 3cm

{\Large {\bf Table Captions}}\\
\vskip 1cm
\begin{description}
\item [Table 1.]  The values of $\Gamma_1^p - \Gamma_1^n$.
The errors are shown only for several points for each set of data.
Uncertanties of our analysis are comparable with ones of 
\cite{SMC,SMCd,E143,Q2E154}.
\item [Table 2.]
The values of the coefficients of CCFR data parametrization.

\end{description}

\newpage
%
%

%



%
%
\hskip -.56cm
{\bf Table 1.}~~ The values of $\Gamma_1^p - \Gamma_1^n$.
The errors are shown only for several points for each set of data.
Uncertanties of our analysis are comparable with ones of 
\cite{SMC,SMCd,E143,Q2E154}.\\
\begin{table}[t]
\begin{center}
\begin{tabular}{|l|c|c|l|l|l|}
\hline
$Q^2$ (GeV$^2$) & 100 & 30 & 10 & 5 & 3 \\
\hline
\multicolumn{6}{|c|}{SMC proton \cite{SMC} and deutron data \cite{SMCd}}\\
\hline
$A_1$-scaling & 0.247 & 0.226 & 0.202 & 0.186 & 0.170 \\
$A_1^*$-scaling & 0.210 & 0.201 & 0.191 & 0.184 & 0.176 \\
Analysis of \cite{SMC} &  &  & 0.183   & 0.181 $\pm$ 0.035 &  \\
\hline
\multicolumn{6}{|c|}{SMC proton \cite{SMC} and E154 neutron data
\cite{Q2E154}}\\
\hline
$A_1$-scaling & 0.221 & 0.209 & 0.194 & 0.183 & 0.171 \\ 
$A_1^*$-scaling & 0.194 & 0.190 & 0.185 & 0.181 & 0.176 \\
\hline
\multicolumn{6}{|c|}{E143 proton and deutron data \cite{E143}}\\
\hline
$A_1$-scaling & 0.170 & 0.169 & 0.165 & 0.160 & 0.154 \\
$A_1^*$-scaling & 0.163 & 0.162 & 0.160 & 0.157 & 0.154 \\
Analysis of \cite{E143} &  &  &   & 0.164 $\pm$ 0.021 & 0.164  \\
\hline
\multicolumn{6}{|c|}{E143 proton \cite{E143} and E154 neutron data
\cite{E154n,Q2E154}}\\
\hline
$A_1$-scaling & 0.189 & 0.186 & 0.179 & 0.174 & 0.169 \\ 
$A_1^*$-scaling & 0.172 & 0.172 & 0.171 & 0.169 & 0.166 \\
Analysis of \cite{Q2E154} &  &  &   & 0.171 $\pm$ 0.013 &    \\
Analysis of \cite{E143} &  &  &   & 0.170 $\pm$ 0.012 &    \\
\hline
Theory & 0.194 & 0.191 & 0.186 & 0.181 $\pm$ 0.002 & 0.177 \\
\hline
\end{tabular}
\end{center}
\end{table}

%
%
\vskip .2cm \hskip -.56cm
{\bf Table 2.}~~The values of the coefficients of CCFR data parametrization.\\
\vskip -.9cm
\begin{table}[h]
\begin{center}
\begin{tabular}{|c|c|c|c|c|c|}
\hline
 $ C_1$ &  $ C_2$ &  $ C_3$ &   $ C_4$ &  $ C_5$ & $ C_6$ \\
\hline
 0.33092 & 3.5000  & 6.5739 & -7.1015 & 2.5388 & 7.6944  \\
\hline
$ C_7$  &  $ C_8$ &  $ C_9$ &   $ C_{10}$ &   $C_{11}$ & \\
\hline
 -8.4285 &  4.9135 &  -4.9857 &  -8.1629 &  1.8193 & \\
\hline
 $ B_1$ &  $ B_2$ &  $ B_3$ &   $ B_4$ &  $ B_5$ & $ B_6$ \\
\hline
 -0.06101 & 3.5000  & 4.9728 & -3.1309 & -1.3361 &  0.94242 \\
\hline
$ B_7$  &  $ B_8$ &  $ B_9$ &   $ B_{10}$ &   $B_{11}$ &
\\
\hline
 0.11729 &  -0.92024 & -1.6489 &  0.61776 &  0.38910 &
\\
\hline
\end{tabular} \end{center} \end{table}


\begin{thebibliography}{40}
%
\bibitem{SMCp} SMC,
 D. Adams, {\it et al.},
Phys. Lett. B {\bf 329}, 399 (1994), B {\bf 339}, 332(E) (1994),
Phys. Lett. B {\bf 357}, 248 (1995);
E143 Collab., K. Abe, {\it et al.},
Phys. Rev. Lett. {\bf 74}, 346 (1995),
Phys. Rev. Lett. {\bf 75}, 25 (1995);
HERMES Collab., K. Ackerstaff, {\it et al.},
Phys. Lett. B {\bf 404}, 383 (1997).
%
\bibitem{SMC} SMC,
 D. Adams, {\it et al.},
Phys. Rev. D {\bf 56}, 5330 (1997).
%
\bibitem{SMCd} SMC,
 D. Adams, {\it et al.},
Phys. Lett. B {\bf 396}, 338 (1997).
%
%
%
\bibitem{E143} E143 Collab., K. Abe, {\it et al.}, SLAC-PUB-7753
(hep-ph/9802357), Phys. Rev. D (in press).
%
\bibitem{E154n} E154 Collab., K. Abe, {\it et al.},
Phys. Rev. Lett. {\bf 79}, 26 (1997).
%
\bibitem{Q2E154}
E154 Collab., K. Abe, {\it et al.}, Phys. Lett. B {\bf 405}, 180 (1997).
%
\bibitem{Gagu}
N. Gagunashvili, {\it et al.}, NIM {\bf 412}, 146 (1998).
%
\bibitem{AEL} M. Anselmino, A. Efremov, E. Leader, Phys. Rep. {\bf
    261} (1995) 1; J. Ellis, M. Karliner, CERN-TH.279/95
and TAUP-2297-95, 1995 (unpublished);
B.L. Ioffe, in {\bf Proceedings of Quarks-94}, Vladimir, p.14.
%
\bibitem{Ram} G.P. Ramsey, Prog. Part. Nucl. Phys. {\bf
    39}, 599 (1997).
%
\bibitem{Bj}
J.D. Bjorken, Phys. Rev. {\bf 148}, 1467 (1966); D {\bf 1}, 1376 (1970).
%
\bibitem{EJ}
J. Ellis,  R.L. Jaffe, Phys. Rev. D {\bf 9}, 1444 (1974); D {\bf 10}, 1669
(1974).
%
\bibitem{EK93} J. Ellis, M. Karliner, Phys. Lett. B {\bf 313}, 131
(1993); F.E. Close, R.G. Roberts, Phys. Lett. B {\bf 316}, 165
(1993).
%
%
\bibitem{Q2A}
 G. Altarelli, P. Nason, G. Ridolfi, Phys. Lett. B {\bf
    320}, 152 (1994);
M. Gluck, E. Reya, W. Vogelsang, Phys. Lett. B {\bf 359}, 201 (1995);
T. Gehrmann, W.J. Stirling, Z. Phys. C {\bf 65}, 470 (1995);
E143 Collab., K. Abe, {\it et al.},
Phys. Lett. B {\bf 364}, 61 (1995); R.D. Ball, S. Forte, G. Ridolfi,
Nucl. Phys. B {\bf 444}, 287 (1995);
A.V. Kotikov, D.V. Peshekhonov,
Phys. Rev. D {\bf 54}, 3162 (1996), Phys. of Atomic Nuclei, {\bf 60}, 653
(1997);
T. Weigl, W. Melnitchouk,
Nucl. Phys. B {\bf 465}, 267 (1996); 
G. Altarelli, {\it et al.}, Nucl. Phys. B {\bf 496}, 337 (1997);
J.P. Nassalski,
 in {\bf Proceeding of the XXVII Internation Conference
on High Energy Physics (1996)} p. 35
(hep-ph/9612352).
%
\bibitem{GRSV}
M. Gluck, E. Reya, M. Stratmann, W.Vogelsang, Phys. Rev. D {\bf 53}, 4775
(1996).
%
%
\bibitem{GS96}
T. Gehrmann, W.J. Stirling, Phys. Rev. D {\bf 53}, 6100 (1996).
%
\bibitem{BFR}
R.D. Ball, S. Forte, G. Ridolfi, Phys. Lett. B {\bf 378}, 225 (1996).
%
\bibitem{HKM}  M. Hirai, S. Kumano, M. Miyama,  
Comp. Phys. Commun. {\bf 108}, 38 (1998).
%
\bibitem{NST}  W.-D. Nowak, A.V. Sidorov, M.V. Tokarev, 
Nuovo Cim. A {\bf 110}, 757 (1997);
E. Leader, A.V. Sidorov, D.B. Stamenov, hep-ph/9708335.
%
%
\bibitem{DIS98}
A.V. Kotikov, D.V. Peshekhonov,
talk on {\bf the Workshop DIS98} (hep-ph/9805374). 
%
\bibitem{Kodaira} J. Kodaira, {\it et al.},
Phys. Rev. D {\bf 20}, 627 (1979); Nucl. Phys. B {\bf 159}, 99 (1979).
%
\bibitem{RoSa} D.A. Ross, C.T. Sachrajda, Nucl. Phys. B {\bf 149}, 497
(1979); G. Altarelli, Phys. Rep. {\bf 81}, 1 (1982).
%
\bibitem{GRam}  M. Goshtasbpour,  G.P. Ramsey,
Phys. Rev. D {\bf 55}, 1244 (1997);
L. E. Gordon, M. Goshtasbpour, G.P. Ramsey, hep-ph/9803351.
%
\bibitem{E581} FNAL
E581/704 Collab., D.L. Adams, {\it et al.}, Phys. Lett. B {\bf 336}, 369
(1994).
%
\bibitem{MeNe} R. Merting, W.L. van Neerven,
 Z. Phys. C {\bf 70}, 625 (1996); W. Vogelsang, Phys. Rev. D
{\bf 54}, 2023 (1996).
%
\bibitem{BoSo} C. Bourrely, J. Soffer,
Nucl. Phys. B {\bf 445}, 341 (1995).
%
\bibitem{Hei} R.L. Heinmann,
Nucl. Phys. B {\bf 64}, 429 (1973).
%
\bibitem{ElKa1} J. Ellis, M. Karliner, Phys. Lett. B {\bf 213}, 73
(1988).
%
\bibitem{BERns}
J. Bartels, B.I. Ermolaev, M.G. Ryskin, Z. Phys. C {\bf 70}, 273 (1996).
%
\bibitem{BERs}
J. Bartels, B.I. Ermolaev, M.G. Ryskin, Z. Phys. C {\bf 72}, 627 (1996).
%
\bibitem{CCFRN} CCFR/NuTeV Collab., W. Seligman,  {\it et al.},
 Phys. Rev. Lett.  {\bf 79}, 1213 (1997);
W. Seligman, Ph.D. thesis, Columbia University, Nevis Report 292, 1997.
%
\bibitem{NMC} NMC, M. Arneodo, {\it et al.},
Phys. Lett. B {\bf 364}, 107 (1995).
%
\bibitem{SLAC} L.W. Whitlow, {\it et al.},
  Phys. Lett. B {\bf 250}, 193 (1990).
%
\bibitem{KP}
A.V. Kotikov, D.V. Peshekhonov,
JETP Lett. {\bf 65}, 7 (1997); in
 {\bf Proceeding of the Workshop DIS96}, p.612
(hep-ph/9608369). 
%
%
\bibitem{Rat}  P. Ratchliffe,
in {\bf Proceeding of the International Symposium SPIN96}, p.161 
(hep-ph/9611348).
%
\bibitem{CCFR} CCFR Collab., P.Z. Quintas,  {\it et al.},
Phys. Rev. Lett. {\bf 71}, 1307 (1993);
M.~Shaevitz, {\it et al.}  Nucl. Phys. Proc. Suppl. B
{\bf 38}, 188 (1995).
%
\bibitem{LRV1}
S.A. Larin, T. van Ritbergen, J.A.M. Vermaseren,
Phys. Lett. B {\bf 404}, 153 (1997).
%
\bibitem{LV}
S.A. Larin, J.A.M. Vermaseren, Phys. Lett. B {\bf 259}, 345 (1991).
%
\bibitem{HNS}
D. van Harrach, W.-D. Nowak, J. Soffer, 
Preprint DESY 97-232, 
CPT-97/P.3564, 1997 (hep-ph/9712207);
A. De Roeck and T. Gehrmann, DESY 97-233, 1997 (hep-ph/9711512).
%
%
%
%
%
\end{thebibliography}
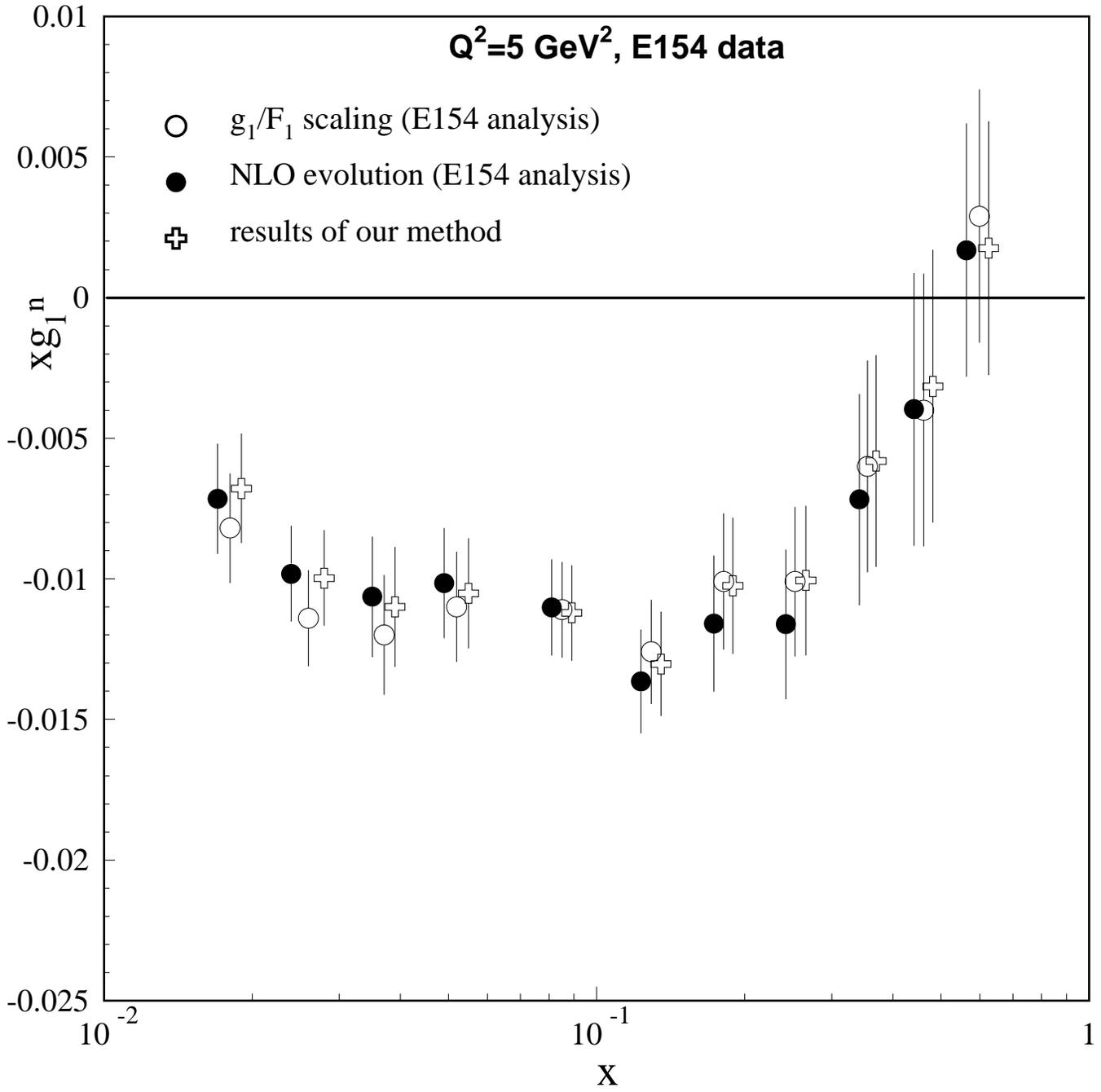
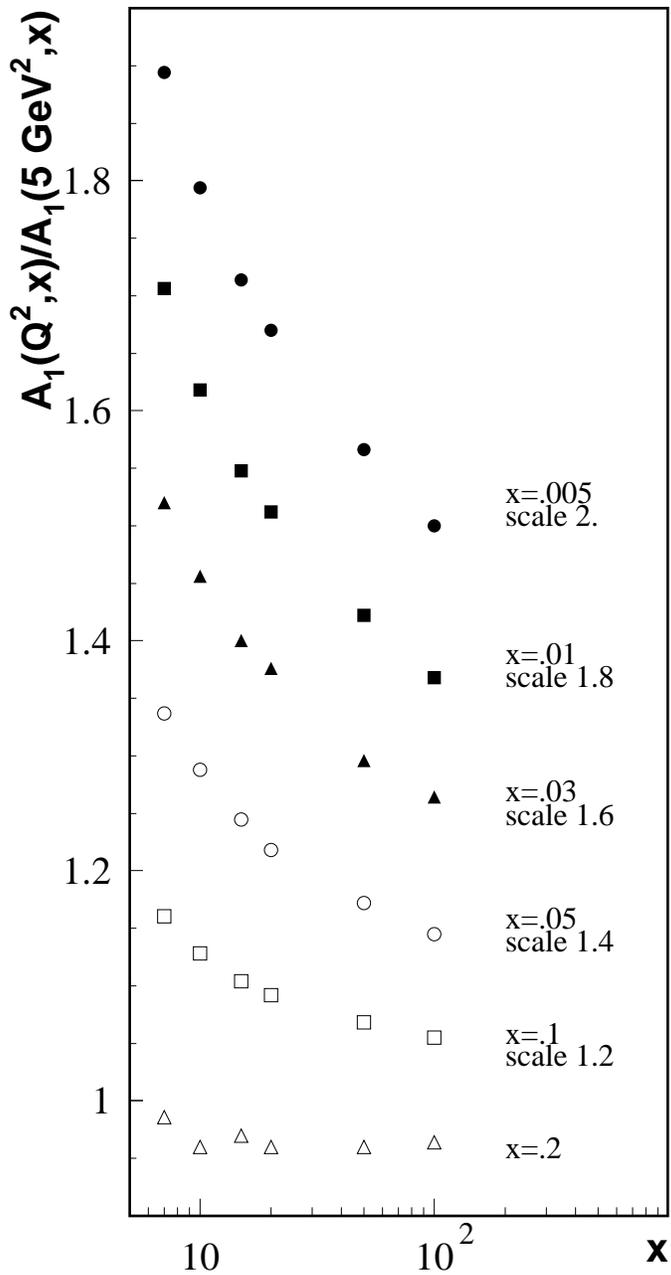
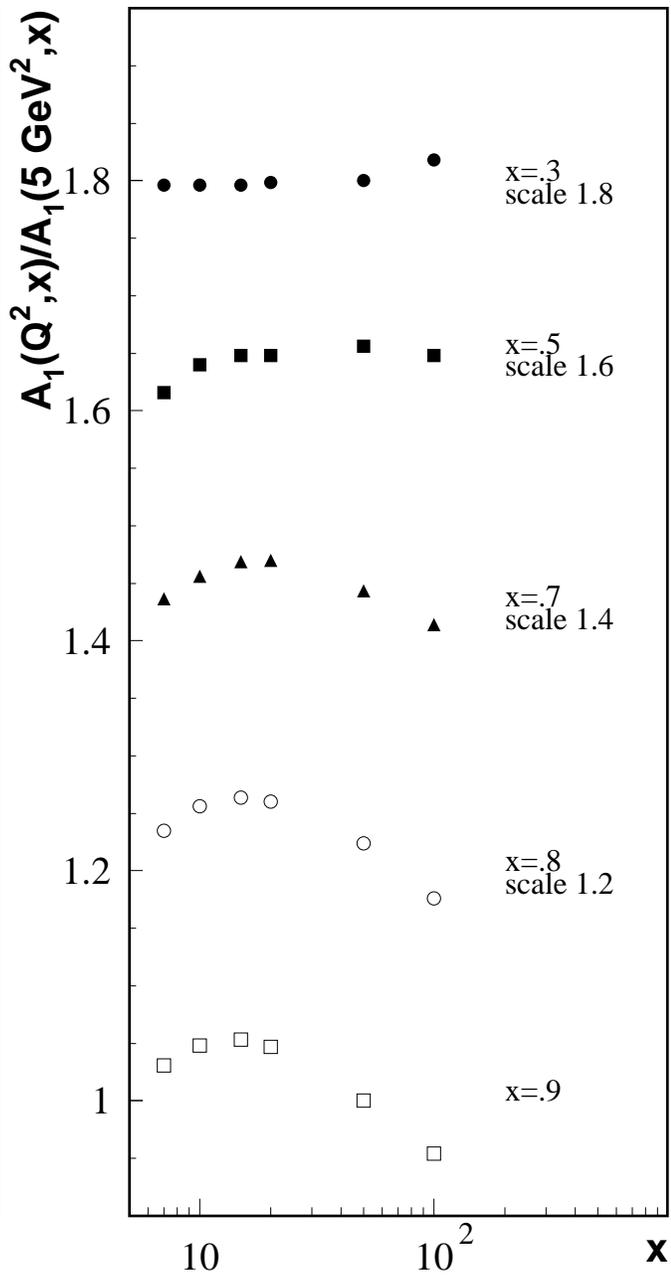
\end{document}